\documentclass{appolb}
\usepackage{epsfig}

\begin{document}
\newcount\eLiNe\eLiNe=\inputlineno\advance\eLiNe by -1
\newcommand{\gsim}{\raisebox{-0.07cm}{$\, \stackrel{>}{{\scriptstyle
\sim}}\, $}}
\title{QCD Analysis of Polarized Scattering Data \\
and \\ 
New Polarized Parton Distributions%
}
\author{Johannes BL\"UMLEIN and Helmut B\"OTTCHER%
\thanks{Talk given at DIS2002, Krakow, Poland, April 30 -- May 4, 2002.}%
\address{DESY Zeuthen, Platanenallee 6, D-15738 Zeuthen, Germany} \\
e-mail: {\tt johannes.blumlein@desy.de} or {\tt hboett@ifh.de}
}
\maketitle

\begin{abstract}
\noindent
In this talk results from a new QCD analysis in leading (LO) and
next--to--leading (NLO) order are presented. New parametrizations of
the polarized quark and gluon densities are derived together with 
parametrizations of their fully correlated 1$\sigma$ error
bands. Furthermore the value of $\alpha_s(M_Z^2)$ is determined.  
Finally a number of low moments of the polarized parton densities are 
compared with results from lattice simulations.   
All details of the analysis are given in Ref.~\cite{jblhfb}.
\end{abstract}
\section{Formalism}
\noindent
The twist--2 contributions to the structure function $g_1(x,Q^2)$ can
be represented in terms of a {\sc Mellin} convolution of the polarized 
singlet density $\Delta \Sigma$, the polarized gluon density $\Delta G$, 
the polarized non--singlet density $\Delta q_j^{NS}$, and the
corresponding polarized Wilson coefficient functions $\Delta C_i^A$
by  
\begin{eqnarray}
\label{eqg1}
g_1(x,Q^2)&=& \frac{1}{2}   \sum_{j=1}^{N_f} e_j^2 \int_{x}^1
{dz \over z} \Bigg [ { 1 \over N_f}\, \Delta \Sigma
\left({x \over z},\mu_f^2\right) \Delta C_q^S
\left(z,{Q^2 \over \mu_f^2}\right) 
 + \Delta G\left({x \over z},\mu_f^2\right)
\nonumber\\
          & &
\times \Delta C^G\left(z,{Q^2 \over \mu_f^2}\right)
+ \Delta   {q_j}^{NS}\left({x \over z},\mu_f^2\right)
\Delta C_q^{NS}\left(z,{Q^2 \over \mu_f^2}\right) \Bigg ]~,
\end{eqnarray}
where $e_j$ denotes the charge of the $j$th quark flavor, $N_f$ 
the number of flavors, and $\mu_f$ is the factorization scale. In
addition the above quantities depend on the renormalization scale
$\mu_r$ of the strong coupling constant $a_s(\mu_r^2) =
g_s^2(\mu_r^2)/(16 \pi^2)$.   

The change of the parton distributions w.r.t. the factorization scale
$\mu_f^2 = Q^2$ is described by the evolution equations which contain
the polarized splitting functions $\Delta P_{ij}$. Both the polarized
Wilson coefficient \cite{coeff} and the polarized splitting functions
\cite{split} are known in the $\overline{ \rm MS}$ scheme up to NLO.

The evolution equations are solved in {\sc Mellin}--$N$ space. A
{\sc Mellin}--transformation turns the {\sc Mellin} convolution into
an ordinary product. The solutions for the evolved parton distributions
are structured such that the input and the evolution part factorize
\cite{evol}, which is of key importance for the error calculation. An
inverse {\sc Mellin}--transformation to $x$--space is then performed
numerically by a contour integral in the complex plane around all
singularities. 
\section{Parametrization and Error Calculation}
\noindent
The shape chosen for the parameterization of the polarized parton
distributions at the input scale of $Q^2 = 4.0~\rm{GeV^2}$ is:

\begin{equation}
x\Delta q_i(x,Q_0^2) = \eta_i A_i x^{a_i} (1 - x)^{b_i} 
(1 + \gamma_i x + \rho_i x^{\frac{1}{2}}).
\end{equation}

\noindent
The normalization constant $A_i$ is defined such that $\eta_i$
is the first moment of $\Delta q_i(x,Q_0^2)$. The densities to be
fitted are $\Delta u_v$, $\Delta d_v$, $\Delta \bar{q}$, and $\Delta
G$.  

Assuming $SU(3)$ flavor symmetry the first moments of $\Delta u_v$ and  
$\Delta d_v$ can be fixed by the $SU(3)$ parameters $F$ and $D$ 
measured in neutron and hyperon $\beta$--decays to 
$\eta_{u_v} = 0.926$ and $\eta_{d_v} = -0.341$. The sea-quark
distribution $\Delta \bar{q}$ was assumed to be described 
according to $SU(3)$ flavor symmetry. Given the present accuracy of
the data we set a number of parameters to zero, namely $\rho_{u_v} =
\rho_{d_v} = 0$, $\gamma_{\bar{q}} = \rho_{\bar{q}} = 0$, and
$\gamma_G = \rho_G = 0$.  
Furthermore we adopted the following two parameter relations: 
$a_G = a_{\bar{q}} + c$, with $0.5 < c < 1.0$ and 
$(b_{\bar{q}}/b_G)(pol) = (b_{\bar{q}}/b_G)(unpol)$. 
These relations were essential to respect positivity for $\Delta
\bar{q}$ and $\Delta G$ and to achieve the expected similar low--$x$
behaviour for both parton densities. No positivity constraint was
assumed for $\Delta u_v$ and $\Delta d_v$. In addition $\Lambda_{QCD}$
was determined.  

The gradients of the polarized parton densities w.r.t. the fitted
parameters needed for the Gaussian error propagation are
calculated analytically at the input scale $Q_0^2$ in {\sc
Mellin}--$N$ space. Their values at $Q^2$ are then given by
evolution. For more details see Ref.~\cite{jblhfb}.

To treat all data sets on the same footing only the statistical errors
were used. To be able to calculate the fully correlated 1$\sigma$
error bands via Gaussian error propagation only fits ending with a
positive--definite covariance matrix were accepted. We allowed for a
relative normalization shift between the different data sets within
the normalization uncertainties quoted by the experiments. Thereby the
main systematic uncertainties of the data were taken into account. The
relative normalization shifts being fitted once and then fixed enter
as an additional term (penalty) for each data set in the
$\chi^2$--expression.   
\section{Results}
\noindent
The results reported here are based on 435 data points of asymmetry
data, i.e. $g_1/F_1$ or $A_1$, above $Q^2 = 1.0~\rm{GeV^2}$, the world
statistics published so far (see Ref. \cite{jblhfb} for a list of
references).  
The fits are performed on $g_1$ which is
evaluated from the asymmetry data using parametrizations for $F_2$
\cite{F2NMC} and $R$ \cite{R1990}. The data do not constrain the four
parameters $\gamma_{u_v}$, $\gamma_{d_v}$, $b_{\bar{q}}$, and $b_G$
well enough. Their values have been fixed as obtained in the
first minimization. The NLO polarized parton densities at the input
scale are presented in Fig. 1 (ISET = 3, see Ref.~\cite{jblhfb}).

\begin{figure}[htb]
\begin{center}
\includegraphics[angle=0, width=10.4cm]{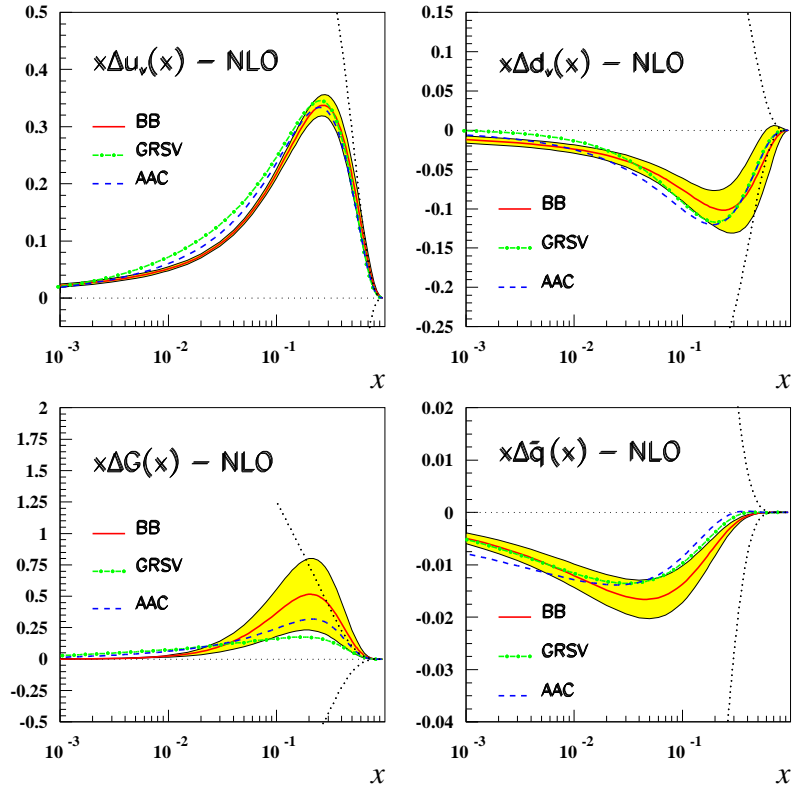}
\end{center}
\vspace*{-0.50cm}
\caption{\label{pdf_comp}
Polarized parton distributions at the input scale $Q_0^2 = 4.0~ GeV^2$ 
(solid line) compared to results obtained by GRSV (dashed--dotted
line) \cite{GRSV}, AAC (dashed line) \cite{AAC}. The shaded areas 
represent the fully correlated $1\sigma$ error bands calculated by 
Gaussian error propagation \cite{jblhfb}. The dark dotted lines
correspond to the positivity bounds given by the unpolarized
densities.} 
\end{figure}

\noindent
The current data constrain $\Delta u_v$ at best, followed by $\Delta
d_v$, $\Delta \bar{q}$, and $\Delta G$. Mainly the lack of data at low
$x$ leads to the broader error bands for the latter two densities.
The comparison with results from other QCD analyses shows the
variation of the outcoming parton distributions when using different
parametrizations at the input scale. The measured structure function
$g_1^p$ is well described both as function of $x$ and $Q^2$. The
derived parton densities and their error bands have been evolved 
to $Q^2$ values up to $10,000~\rm{GeV^2}$. One finds that $\Delta u_v$
stays positive, $\Delta d_v$ remains negative although less constraint
than $\Delta u_v$, and $\Delta G$ stays positive even within the error
band for the range $Q^2 \geq 4\,GeV^2$. All three densities evolve
towards smaller values of $x$. The sea-quark density $\Delta \bar{q}$
is negative for $Q^2 \leq 10^4\,GeV^2$ and remains negative within
errors for $x \leq 5 \cdot 10^{-2}$ up to $Q^2 = 10^4\,GeV^2$, but
changes sign for larger values of $x$.

In determining $\alpha_s$ the QCD--parameter $\Lambda_{QCD}$ was
fitted. The impact of the variation of both the renormalization and
the factorization scale on the value of $\alpha_s$ was investigated. 
The following result was obtained (for ISET = 3, see
Ref.~\cite{jblhfb}, as well for more details):
\vspace*{-0.10cm}
\begin{eqnarray}
\alpha_s(M_Z^2) = 0.113 \begin{array}{c} + 0.004 \\ - 0.004
\end{array}  
{\rm (stat)}
\begin{array}{c} + 0.004 \\ - 0.004 \end{array}{\rm (fac)}
\begin{array}{c} + 0.008 \\ - 0.005 \end{array}{\rm (ren)}~.
\nonumber 
\end{eqnarray}
\noindent
This value is compatible within 1$\sigma$ with the world average of
$0.118 \pm 0.002$ \cite{PDG}, although our central value is somewhat
lower. It is also compatible with results from other QCD analyses 
of polarized and unpolarized data.

In recent lattice simulations \cite{lattice} low moments for the
polarized parton densities $\Delta u_v$, $\Delta d_v$, and $\Delta u -
\Delta d$ were determined. In Table 1 these moments are compared with
the ones extracted from our NLO polarized densities.
\renewcommand{\arraystretch}{1.2}
\begin{center}
\begin{tabular}{|c|r|r|r|r|}
\hline 
\hline 
\multicolumn{1}{|l|}{ }&
\multicolumn{1}{c|}{ }&
\multicolumn{1}{c|}{\tt QCD results }&
\multicolumn{2}{c|}{ lattice results } \\
\cline{3-5}
\multicolumn{1}{|c|}{ $\Delta f$ }& 
\multicolumn{1}{ c|}{ $n$ } &
\multicolumn{1}{ c|}{NLO moments} &
\multicolumn{1}{ c|}{QCDSF    } &
\multicolumn{1}{ c|}{LHPC/    } \\
\multicolumn{1}{|c|}{ }&
\multicolumn{1}{ c|}{   } &
\multicolumn{1}{ c|}{ at $Q^2=4~\rm{GeV^2}$ } &
\multicolumn{1}{ c|}{         } &
\multicolumn{1}{ c|}{SESAM    } \\
\hline \hline
$\Delta u_v$  &--1 & $0.926 \pm 0.071$ & 0.889(29) & 0.860(69) \\
              &  0 & $0.163 \pm 0.014$ & 0.198(8) & 0.242(22) \\
              &  1 & $0.055 \pm 0.006$ & 0.041(9) & 0.116(42) \\
\hline
$\Delta d_v$  &--1 & $-0.341 \pm 0.123$ & -0.236(27) & -0.171(43) \\
              &  0 & $-0.047 \pm 0.021$ & -0.048(3) & -0.029(13) \\
              &  1 & $-0.015 \pm 0.009$ & -0.028(2) &  0.001(25) \\
\hline
$\Delta u$--$\Delta d$
              &--1 & $1.267  \pm 0.142$ &  1.14(3) &  1.031(81) \\
              &  0 & $0.210  \pm 0.025$ &  0.246(9) &  0.271(25) \\
              &  1 & $0.070 \pm 0.011 $ &  0.069(9) &  0.115(49) \\
\hline \hline
\end{tabular}
\end{center}
\renewcommand{\arraystretch}{1}
\vspace*{+0.20cm}
{\sf Table 1: Moments of the NLO parton densities (for ISET =3, see
\cite{jblhfb}) at $Q^2 = 4~\rm{GeV^2}$ and from recent lattice
simulations at the scale $\mu^2 = 1/a^2 \sim 4~\rm{GeV^2}$.}

\noindent
The first moment is denoted with $n = -1$. For the $n = 0,1$ values of
the QCDSF Collaboration no continuum extrapolation was performed. 
The values compare within errors. Unlike in the unpolarized case where
a strong $m_{\pi}$--dependence is expected in the lattice extrapolation
the comparison here suggests a flat behaviour, which also has been
found in \cite{DMT} very recently. 
\section{Conclusions}
\noindent
A QCD analysis in LO and NLO of the current world--data on polarized
structure functions was performed. New parametrizations of the
polarized parton densities including their fully correlated 1$\sigma$
error bands were derived.These parameterizations are available as a
fast {\tt FORTRAN}--routine which makes their application possible
in Monte Carlo simulations. The value determined for $\alpha_s(M_Z^2)$
is compatible within 1$\sigma$ with the world average. Comparing the
lowest moments with values from lattice simulations the errors
improved during recent years and the values became closer.
  
\vspace*{+0.25cm}
\noindent
{\bf Acknowledgment}.~This work was supported in part by EU contract
FMRX-CT98-0194 (DG 12 - MIHT). 
%

%
%

\begin{thebibliography}{99}
%
\bibitem{jblhfb}
J.~Bl\"umlein and H.~B\"ottcher, hep-ph/0203155, Nucl. Phys. {\bf B}
in print; 
\bibitem{coeff}
W.L. van Neerven and E.B. Zijlstra, Nucl. Phys. {\bf B417} (1994) 61,
Nucl. Phys. {\bf B426} (1994) 245. 
\bibitem{split}
R. Mertig and W.L. van Neerven, Z. Phys. {\bf C70} (1996) 637,\\
W. Vogelsang, Phys. Rev. {\bf D54} (1996) 2023.
\bibitem{evol}
W.~Furmanski and R.~Petronzio, Z. Phys. {\bf C11} (1982) 283;
M.~Gl\"uck, E.~Reya, and A.~Vogt, Z. Phys. {\bf C48} (1990) 471; 
J.~Bl\"umlein and A.~Vogt, Phys. Rev. {\bf D58} (1998) 014020.
\bibitem{F2NMC}
M.~Arneodo et al., Phys. Lett. {\bf B364} (1995) 107.
\bibitem{R1990}
L.~Withlow et al., Phys. Lett. {\bf B250} (1990) 193.
\bibitem{GRSV}
M.~Gl\"uck et al., Phys. Rev. {\bf D63} (2001) 094005.
\bibitem{AAC}
Y.~Goto et al., Phys. Rev. {\bf D62} (2000) 034017. 
\bibitem{PDG}
Particle Data Group, Eur. Phys. J. {\bf C15} (2000) 91. 
\bibitem{lattice}
M.~G\"ockeler et al., QCDSF Coll., Phys.Rev. {\bf D53} (1996) 2317;
Phys.Lett. {\bf B414} (1997) 340; hep-ph/9711245; Phys.Rev. {\bf D63}
(2001) 074506; S.Capitani et al., Nucl.Phys.(Proc. Suppl.) {\bf
B79} (1999) 548; 
S~.G\"usken et al., SESAM Coll., hep-lat/9901009; 
D~.Dolgov et al., LHPC and SESAM Coll., hep-lat/0201021.
\bibitem{DMT}
M.~Detmold, W.~Melnitchouk, and A.W.~Thomas, hep-lat/0206001.
%
\end{thebibliography}
\end{document}